\documentclass[aps,pra,onecolumn,showpacs,nofootinbib,notitlepage]{revtex4-1}
\usepackage{amsmath,epsfig}
\usepackage{mathtools,fullpage}

\newcommand{\fgies}[3]{\mbox{\raisebox{#3}
{\epsfig{file=#1,scale=#2,clip=true}}~}}
\newcommand{\ph}{\phantom{0}}
\newcommand{\bml}{\begin{multline}}
\newcommand{\bea}{\begin{eqnarray}}
\newcommand{\eea}{\end{eqnarray}}
\newcommand{\be}{\begin{equation}}
\newcommand{\ee}{\end{equation}}
\newcommand{\bi}{\begin{itemize}}
\newcommand{\ei}{\end{itemize}}

\newcommand{\ra}{\rangle}
\newcommand{\la}{\langle}

\newcommand{\up}{\uparrow}
\newcommand{\down}{\downarrow}

\newcommand{\eq}{\,\hat{=}\,}

\newcommand{\bcbp}{\begin{comment}}
\newcommand{\ecbp}{\end{comment}}
\begin{document}

\title{Skeleton series and
multivaluedness
of the self-energy functional
\\in zero space-time dimensions}
\author{Riccardo Rossi$^1$ and F\'elix Werner$^2$}
\affiliation{Laboratoire de Physique Statistique$\,^1$ and Laboratoire Kastler Brossel$\,^2$, Ecole Normale Sup\'erieure, 
UPMC, Universit\'e Paris Diderot$\,^1$,
Coll\`ege de France$\,^2$,
CNRS,
24 rue Lhomond,
 75005 Paris, France}

\begin{abstract}
%In the recent work [E. Kozik, M. Ferrero and A. Georges, PRL {\bf 114}, 156402 (2015)],
% it was found numerically
Recently,
Kozik, Ferrero and Georges have discovered numerically
that for a family of fundamental models of interacting fermions,
the self-energy $\Sigma[G]$ is a multi-valued functional of the fully dressed single-particle propagator $G$,
and that the skeleton diagrammatic series
$\Sigma_{\rm bold}[G]$
 converges to the wrong branch above a critical interaction strength.
We consider
the zero space-time dimensional case,
where the same mathematical phenomena appear from elementary algebra.
%point out that these mathematical phenomena also occur for
 %an exactly solvable fermionic 
%a toy-model in zero space-time dimensions.
We also find a similar phenomenology
for
the fully bold formalism built on fully dressed single-particle propagator and pair propagator.
\end{abstract}

\maketitle

In quantum many-body physics, an important role is played by
the self-consistent field-theoretical %framework
formalism, where the self-energy $\Sigma$ is expressed in terms of the exact propagator $G$ 
(see, {\it e.g.},~\cite{HaussmannBook} and Refs. therein).
In a recent article,
Kozik, Ferrero and Georges numerically
discovered mathematical difficulties with
%  the standard self-consistent many-body theoretical 
this %framework
formalism~\cite{Kozik_2_solutions}.
They studied not only the Hubbard model in two space dimensions,
but also simpler models
---the Hubbard atom and the Anderson impurity model---
for which there is no spatial coordinate.
%In the present note,
Here
we consider an even simpler toy-model
for which there is no imaginary-time coordinate either.
%Similar strategy was followed in
This idea was also followed in
the very recent article~\cite{reining_2solutions}.
%$[\varphi_\sigma(\rr,\tau), \bar{\varphi}_\sigma(\rr,\tau)]$, where
%$\sigma\in\{\uparrow,\downarrow\}$ is the spin index,
%$\rr$ is a $d$-dimensional space coordinate,
%and is $\tau$ the imaginary time.
For fermionic many-body problems,
%, such as the ones studied by Kozik {\it et al.}~\cite{Kozik_2_solutions},
the partition function can be written as a functional integral over Grassmann fields 
in $(d+1)$ space-time dimensions
($d$ spatial coordinates and one imaginary-time coordinate)~\cite{NegeleOrland}.
Accordingly,
we consider the zero space-time dimensional model defined
by a ``partition function'' given by a simple Grassmann integral,
\begin{equation}
%Z(\mu, U) 
Z= \int \left(\prod_{\sigma} d\varphi_\sigma d\bar{\varphi}_\sigma\right)\; e^{-S[\bar{\varphi}_\sigma, \varphi_\sigma]}
\label{eq:def_Z}
\end{equation}
with the action
\be
S[\bar{\varphi}_\sigma, \varphi_\sigma] = -\mu \sum_\sigma  \bar{\varphi}_\sigma  \varphi_\sigma + U \bar{\varphi}_\uparrow \varphi_\uparrow \bar{\varphi}_\downarrow \varphi_\downarrow,
\ee
and a corresponding propagator
\be
G = \la \bar{\varphi}_\sigma
\varphi_\sigma \ra
=
\frac{1}{Z}\ \int \left(\prod_\sigma d\varphi_\sigma d\bar{\varphi}_\sigma\right)\; e^{-S[\bar{\varphi}_\sigma, \varphi_\sigma]} \ \bar{\varphi}_\sigma
\varphi_\sigma.
\label{eq:def_G}
\ee
Here $\sigma\in\{\uparrow,\downarrow\}$ is the spin index,
while
$\mu$ and $U$
are dimensionless parameters that
play the roles of chemical potential and interaction strength.
%Note that the Feynman diagram topologies of the present toy model are identical to the ones of the $(d+1)$ dimensional case.
Diagrammatically, the Feynman rules for the present toy-model are analogous 
to the ones of the physical $(d+1)$ dimensional models,
with the simplification that space-time variables are absent and that the propagators are constants.

%the bare propagator $G_0$ is space-time independent.

In this exactly solvable toy-model,
we observe a similar phenomenology than
the one found by
  Kozik {\it et al.}
in non-zero space-time dimensions.
More precisely,
restricting to $U<0$,
 we find that:
\bi
\item The mapping $G_0 \mapsto G(G_0,U)$ is two-to-one
and hence the function $G\mapsto \Sigma(G,U)$ has two branches.
\item
The skeleton series $\Sigma_{\rm bold}(G,U)$,
evaluated at the exact $G(\mu,U)$,
converges to the correct branch for ${|U|<\mu^2}$, and to the wrong branch for  $|U|>\mu^2$.
\ei

This can be derived very directly from the above definitions.
Expanding the exponentials in Eqs.~(\ref{eq:def_Z},\ref{eq:def_G}) yields
\be
Z(\mu,U)=\mu^2 - U
\ee
\be
G(\mu,U) = \frac{\mu}{\mu^2-U}.
\label{eq:G_vs_mu}
\ee
%({\it e.g.} using $\partial Z/\partial \mu = 2\,Z\,G$).
The  propagator for $U=0$ is
\be
G_0(\mu) = \frac{1}{\mu}.
\label{eq:G0_vs_mu}
\ee
The self-energy $\Sigma$, defined as usual by the Dyson equation
$G^{-1} = G_0^{-1} - \Sigma$,
reads
\be
\Sigma(\mu,U) = \frac{U}{\mu}.
\label{eq:Sig=}
\ee

We note that for $U>0$, an obvious pathology appears in this model around $U = \mu^2$;
namely, $Z$ changes sign, and $G$ diverges.
Therefore we restrict to $U<0$.

Eliminating $\mu$ between Eqs.~(\ref{eq:G_vs_mu},\ref{eq:G0_vs_mu})
gives
\be
G(G_0,U) = \frac{G_0}{1-U G_0^2}.
\label{eq:G(G0)}
\ee
The map $G_0 \mapsto G(G_0,U)$ is two-to-one,
because the $G_0$'s that correspond to a given $G$ are the solutions of the second order equation
\be
U G\, G_0^2 + G_0 - G = 0,
\ee
which has the two solutions
\be
G_0^{(\pm)}(G,U) = \frac{-1 \pm \sqrt{1 + 4\, U G^2}}{2\, U G}.
\label{eq:G0+-(G)}
\ee
These solutions are real provided
$(G,U)$ belongs to the physical manifold
$\{(G(\mu,U),U)\}$;
indeed,
\be
4|U|G(\mu,U)^2\leq 1.
\label{eq:<=1}
\ee
%These solutions are real, since one can check from Eq.~(\ref{eq:G_vs_mu}) that $1 + 4\, U G^2>0$. Thus, for a fixed $U$, and for any $G$ belonging to the set of allowed values for this $U$, there are two  values of $G_0$ ({\it i.e.} of $\mu$)...
The corresponding self-energies (given by the Dyson equation) are
\be
\Sigma^{(\pm)}(G,U) = \frac{-1 \pm \sqrt{1 + 4\, U G^2}}{2\,G}.
\label{eq:Sig_pm}
\ee
The correct  self-energy
$\Sigma(\mu,U)$
is recovered from
$\Sigma^{(s)}\!\left(G(\mu,U),U\right)$
provided one takes the determination
\be
s={\rm sign}(\mu^2-|U|).
\ee

We turn to a discussion of the skeleton diagrammatic series $\Sigma_{\rm bold}(G,U)$ for the self-energy $\Sigma$ in terms of fully dressed propagator $G$ and bare vertex $U$.
We find that 
$\Sigma_{\rm bold}(G,U)$ is
 the $U{\to}0$ Taylor series  of
$\Sigma^{(+)}(G,U)$.
%, and  this series converges to $\Sigma^{(+)}(G,U)$, except at $|U|=\mu^2$ where it diverges.
%%For a complete proof, see the Appendix.
% the slightly cumbersome but straightforward proof of this fact, we just mention that clearly,
Before deriving this, we note that obviously,
$\Sigma_{\rm bold}(G,U)$ can never be 
the Taylor series of
$\Sigma^{(-)}(G,U)$, since the former vanishes at $U=0$  while the latter does not.
%To show that $\Sigma_{\rm bold}$ is given by $\Sigma^{(+)}$,
For the derivation,
 it is convenient to introduce 
$g := \sqrt{|U|}\,G$
and
$g_0 := \sqrt{|U|}\,G_0$, so that Eqs.(\ref{eq:G(G0)},\ref{eq:G0+-(G)}) simplify to
$g(g_0)=g_0/(1+g_0^{\ph 2})$
and
$g_0^{(\pm)}(g) = (1\mp\sqrt{1-4g^2})/(2g)$.
The key point is that
$g_0^{(+)}(g(g_0)) \eq g_0$,
where the symbol $\eq$ means equality
in the sense of formal power series.
This is because 
the inverse mapping of $g(g_0)$ is
$g_0^{(+)}(g)$ for small $g_0$ and $g$.
Let us then denote by $\Sigma_{\rm bare}(G_0,U)$ the diagrammatic series for the self-energy in terms of bare propagators and vertices.
Setting $\Sigma_{\rm bold}(G,U) =: \sqrt{|U|} \, \sigma_{\rm bold}(g)$ and
$\Sigma_{\rm bare}(G_0,U) =: \sqrt{|U|} \, \sigma_{\rm bare}(g_0)$,
a defining property of $\sigma_{\rm bold}$ is that
$\sigma_{\rm bold}(g(g_0)) \eq  \sigma_{\rm bare}(g_0)$.
%, {\it i.e.}, $\sigma_{\rm bold}(g) \eq  \sigma_{\rm bare}(g_0^{(+)}(g))$.
In the present toy-model, we simply have $\Sigma_{\rm bare}(G_0,U) = U\,G_0$, {\it i.e.},
$\sigma_{\rm bare}(g_0) = -g_0$.
Hence, $\sigma_{\rm bold}(g)\eq-g_0^{(+)}(g)$,
{\it i.e.},
$\Sigma_{\rm bold}(G,U) \eq U\,G_0^{(+)}(G,U) = \Sigma^{(+)}(G,U)$.

Explicitly, 
expanding the square root in Eq.~(\ref{eq:Sig_pm}) yields
\be
\Sigma_{\rm bold}(G,U) = \sum_{n=1}^\infty \frac{(-1)^{n-1}\,(2n-2)!}{n!\,(n-1)!}\,G^{2n-1}U^n.
\label{eq:Sig_bold}
\ee

It is natural to evaluate the bold series at the exact $G(\mu,U)$.
The obtained series always converges,
as follows from
the inequality~(\ref{eq:<=1}).
%and from Stirling's formula.
The convergence is always to $\Sigma^{(+)}(G(\mu,U),U)$,
which as we have seen is the correct result for $|U|<\mu^2$,
and the wrong one for $|U|>\mu^2$.
The convergence speed
is
slow for $|U|$ close to $\mu^2$,
and
gets faster
not only in the small~$|U|$ limit,
but also
 in the large $|U|$ limit.
This is qualitatively identical to
the numerical observations of Kozik {\it et al.}
in non-zero space-time dimensions. 
We note that the series converges even
at the critical value $|U|=\mu^2$, 
albeit very slowly (the summand behaving as $1/n^{3/2}$ for large $n$);
at this point, the boundary of the series' convergence disc is reached.

In the Figure we plot the quantity $\Sigma\,G$,
which,
for the exact $\Sigma$,
 is equal to $U$ times the double occupancy $\la \bar{\varphi}_\up \varphi_\up \bar{\varphi}_\down \varphi_\down \ra$, 
versus $|U|$ for fixed $\mu$.
The picture is qualitatively identical to Fig.~2(a) of Kozik~{\it et~al.}

\begin{figure}
\includegraphics{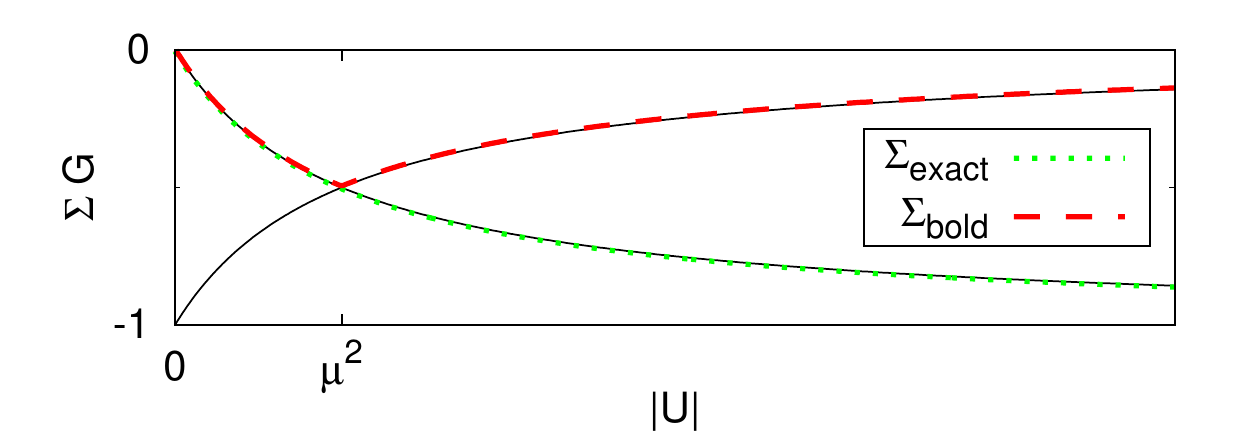}
\caption{The two branches of the self-energy $\Sigma$,
multiplied for convenience by $G$, as a function of the interaction strength $|U|$, for fixed $\mu$.
Dotted line: exact self-energy $\Sigma_{\rm exact}=\Sigma(\mu,U)$,
dashed line: skeleton series $\Sigma_{\rm bold}(G,U)$.
The upper branch corresponds to $\Sigma^{(+)}(G,U)$, the lower branch to $\Sigma^{(-)}(G,U)$.
Here, $G$ stands for the exact $G(\mu,U)$.
}
\end{figure}

Geometrically, the mapping $U\mapsto \Sigma(G,U)$ can be viewed as single-valued on a two-sheeted Riemann-surface
with a branch point at $-1/(4G^2)$.
Let us vary $U$ from $0$ to $-\infty$ for fixed $\mu$.
For small $|U|$,
the point $U$ is far away from the branch point
and
the bold series
converges quickly.
The result corresponds to the correct Riemann-sheet.
Upon increasing $|U|$, the point $U$ and the branch point $-1/(4G^2(\mu,U))$ both move leftwards.
The point $U$ 
catches up the branch point when $|U|=\mu^2$. For larger $|U|$, $U$ is again to the right of the branch point,
and the bold series converges again,
 but the result corresponds to the wrong sheet.
In principle, the correct result can be recovered from
$\Sigma_{\rm bold}(G,U)$
by analytic continuation along a path where $U$ rotates once around the branch point.

As emphasized by Kozik {\it et al.}, since
the self-energy is the functional derivative of the Luttinger-Ward functional $\Phi[G,U]$ with respect to $G$,
and since $\Sigma[G,U]$ is multivalued, 
$\Phi$ must also be multivalued.
This can also be seen explicitly in the present model.
The Luttiger-Ward functional (which is actually a function in the present model) can be constructed following the usual procedure (see, {\it e.g.},~\cite{HaussmannBook}).
Starting from the free energy $F(\mu,U)=-\ln Z(\mu,U)$,
and noting that
\be
\frac{\partial F}{\partial \mu}(\mu,U) = -2\,G,
\label{eq:F'=G}
\ee
the Baym-Kadanoff functional is defined by Legendre transformation:
\be
\Omega(G,U) = F(\mu,U) + 2 \, \mu\, G
\label{eq:Omeg}
\ee
with $\mu(G,U)$ such that Eq.~(\ref{eq:F'=G}) holds.
The Luttinger-Ward functional is then defined by
\be
\Phi(G,U) = \Omega(G,U) - \Omega(G,0).
\ee
This leads to the expression
\be
\Phi_{\pm}(G,U) = -\ln|\Sigma^{(\mp)}(G,U)| - \ln|G| - 2\, G\, \Sigma^{(\mp)}(G,U) - 2.
\ee
There are two branches because Eq.~(\ref{eq:F'=G}) has two solutions.
Accordingly, the mapping $\mu\mapsto F(\mu,U)$ is neither convex nor concave.
Finally one can check that\footnote{The unconventional factors $2$ in Eqs.~(\ref{eq:F'=G},\ref{eq:Omeg},\ref{eq:dPhi}) could be removed by working with spin-dependent $\mu$ and $G$.}
\be
\frac{\partial \Phi_{\pm}}{\partial G}(G,U) = 2\, \Sigma^{(\pm)}(G,U).
\label{eq:dPhi}
\ee

%It remains unclear whether the present complete understanding of the zero space-time dimensional case will improve the understanding of the higehr-dimensional cases.
We point out
that 
in the present zero space-time dimensional model,
both branches $G_0^{(\pm)}(G,U)$ are physical in the sense that 
they are the non-interacting propagator for certain parameters of the model.
%there exist model-parameters $\mu_{\pm}$ such that $G_0^{(\pm)}=G_0(\mu_{\pm})$.
This is not the case for the  Hubbard~atom and the Hubbard~model~\cite{Kozik_2_solutions}.
More generally, the absence of imaginary-time coordinate consitutes a drastic simplification, and while we have observed similar phenomena than in~\cite{Kozik_2_solutions}, 
it remains an open question
to which extent the underlying mechanisms are similar.

Finally we briefly treat the fully bold formalism built not only on the fully dressed $G$ but also on the fully dressed pair propagator $\Gamma$.
This formalism was used, {\it e.g.}, in Refs.~\cite{Reson2papers,DengEmergentBCS}.
One defines
\be
\Gamma = U - U^2 \la \varphi_\down \varphi_\up \bar{\varphi}_\up \bar{\varphi}_\down \ra
\ee
or diagrammatically
\be
%{\hskip -2cm} 
\fgies{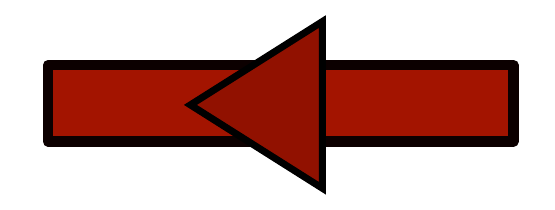}{0.23}{-1.3mm} 
= \,  \fgies{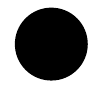}{0.23}{-.5mm} \, + \,   \fgies{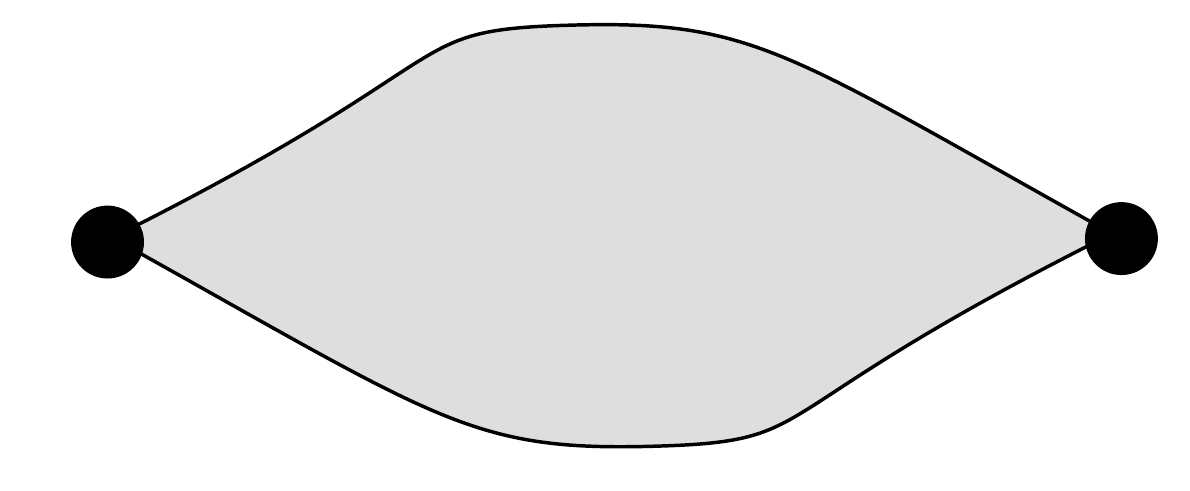}{0.23}{-5mm}.
\ee
The pair self-energy $\Pi$ is defined by the Dyson equation
$\Gamma^{-1} = U^{-1} - \Pi$.
\\
The dressed  $G$ and $\Gamma$ are given in terms of the bare $G_0$ and $U$ by Eq.~(\ref{eq:G(G0)}) and
\be
\Gamma = U \ \frac{1 - 2\,U\,G_0^{\ph 2}}{1 - U\,G_0^{\ph 2}}.
\ee
Eliminating $U$ yields a cubic equation for $G_0$, which reads
\begin{equation}\label{cubique}
\gamma g_0^3+2g_0^2-3g_0+1=0
\end{equation}
in terms of the rescaled quantities
\bea
g_0 &:=& G_0/G
\\
 \gamma &:=& \Gamma \,G^2.
\eea
In the relevant range $0<|\gamma|\leq 2\sqrt{3}/9$ the three solutions are
\be
g_0^{(l)}=\frac{2}{3 \gamma }\left[ \sqrt{9\gamma+4}\;\cos{\left(\frac{\theta(\gamma)+2\pi l}{3}\right)} - 1\right],\ \ \ \ \ \ \ \  l\in\{-1,0,1\} 
\ee
where $\theta(\gamma)=\arg(-27\gamma^2-54\gamma-16-3i\sqrt{3}|\gamma|\sqrt{4-27\gamma^2})\in(-\pi,\pi)$.
The fully bold diagrammatic series $\Sigma_{\rm bold}(G,\Gamma)$ and $\Pi_{\rm bold}(G,\Gamma)$, evaluated at the exact $G(\mu,U)$ and $\Gamma(\mu,U)$,
always converge to the $l=1$ branch\footnote{Explicitly, we have
$\Sigma_{\rm bold}(G,\Gamma) \, \hat{=} \, [1/g_0^{(+1)}(\gamma)-1]/G$
and
$\Pi_{\rm bold}(G,\Gamma) \, \hat{=} \, \Sigma_{\rm bold}(G,\Gamma) / [G\,g_0^{(+1)}(\gamma)]$.}, which coincides with the exact result for $|U|/\mu^2<(1+\sqrt{3})/2$. Above this critical interaction strength, the exact result is the $l=-1$ branch, so that the bold series converges to a wrong result. 
At this critical interaction, the boundary of the series' convergence disc is reached.
In summary, a similar phenomenology occurs again.
%However, there is no reason to expect that this affects the resonant gas study of Refs.~\cite{Reson2papers}, given that the problem was only observed at or close to half filling in Ref.~\cite{Kozik_2_solutions}.

%A detailed study of
The consequences
for the 
Bold Diagrammatic Monte Carlo approach~\cite{SvistunovProkofevPolaronBold2papers,Reson2papers,Kulagin2papers,MishchenkoProkofevPRL2014,DengEmergentBCS}
of the findings of Ref.~\cite{Kozik_2_solutions} and of the present work
is left for future study.

We thank E.~Kozik, N.~Prokof'ev and K.~Van~Houcke for discussions.
We acknowledge support from CNRS, ERC and IFRAF-NanoK
(grants PICS 06220, `Thermodynamix',
and
`Atomix').
Our host institutions are members of `Paris~Sciences~et~Lettres', `Sorbonne~Universit\'es', and `Sorbonne~Paris~Cit\'e'.

%{\it Acknowledgements:}

%\appendix
%\section{Proof of $\Sigma_{\rm bold}=\Sigma^{(+)}$}

\bibliography{felix_copy}
\end{document}